\documentclass[preprint2]{aastex61}

\usepackage{amsmath}	
\usepackage{amssymb}	
\usepackage{wasysym}

\newcommand\aastex{AAS\TeX}

\newcommand{\sitem}{\vspace{-0.3cm}\item}

\newcommand{\fHHe}{f_{\text{H-He}}}
\newcommand{\MHHe}{M_{\text{H-He}}}
\newcommand{\fwater}{f_{\text{H$_2$O}}}
\newcommand{\Mwater}{M_{\text{H$_2$O}}}
\newcommand{\Mrock}{M_{\text{rock}}}
\newcommand{\frock}{f_{\text{rock}}}

\newcommand{\RE}{R$_{\oplus}$}

\received{--}
\revised{--}
\accepted{\today}

\submitjournal{ApJ}

\shorttitle{\aastex\ Exoplanets thresholds}
\shortauthors{Lozovsky et al.}

\begin{document}

\title{Threshold  radii of  volatile-rich planets} 

\correspondingauthor{Michael Lozovsky}
\email{michloz@mail.com}

\author{M. Lozovsky}
\affiliation{Center for Theoretical Astrophysics \& Cosmology, 
Institute for Computational Science, 
University of Zurich, Zurich, Switzerland}
\author{R. Helled}
\affiliation{Center for Theoretical Astrophysics \& Cosmology, 
Institute for Computational Science, 
University of Zurich, Zurich, Switzerland}
\author{C. Dorn}
\affiliation{Center for Theoretical Astrophysics \& Cosmology, 
Institute for Computational Science, 
University of Zurich, Zurich, Switzerland}
\author{J. Venturini}
\affiliation{Center for Theoretical Astrophysics \& Cosmology, 
Institute for Computational Science, 
University of Zurich, Zurich, Switzerland}

\begin{abstract}
Constraining the planetary composition is essential for exoplanetary characterization. In this paper, we use a statistical analysis to determine the characteristic maximum (threshold) radii for various compositions for exoplanets with masses up to 25 Earth masses (M$_\oplus$). 
We confirm that  most planets with radii larger than 1.6 Earth radius (R$_\oplus$) are not rocky, and must consist of lighter elements, as found by previous studies. 
We find that planets with radii above 2.6 R$_\oplus$ cannot be pure-water worlds, and must contain significant amounts of hydrogen and helium (H-He). We find that planets with radii larger than about 3 R$_\oplus$, 3.6 R$_\oplus$, and 4.3 R$_\oplus$ are expected to consist of  2\%, 5\% and 10\% of H-He,  respectively. 
We investigate the sensitivity of the results to the assumed internal structure, the planetary temperature and albedo, and the accuracy of the mass and radius determination. 
We show that the envelope's metallicity, the percentage of H-He and the distribution of the elements play a significant role in the determination of the threshold radius. Finally, we conclude that despite the degenerate nature of the problem, it is possible to put limits on the possible range of compositions for planets with well-measured mass and radius.   

\end{abstract}

\keywords{exoplanets -- planets and satellites: detection -- planets and satellites: fundamental parameters}

\section{Introduction} \label{sec:intro}
The ongoing efforts to detect and characterize exoplanets from Earth and space have led to the detection of thousands of exoplanets, and allows us to study planets as a class of astrophysical objects. Measured radii of planets from the {\it Kepler} mission combined with Radial Velocity (RV) and Transit Timing Variations (TTV) follow-ups provide information on the planetary radii and masses, and therefore, on their mean densities. 
The measured masses and radii can be compared to theoretical mass-radius (M-R) relations of planetary objects, which is used to infer the possible bulk composition \citep[e.g.][]{Weiss2014,Zeng2016,Wolfgang2015,Batygin2013}. 
\par 

Since the discovery of exoplanets with radii between that of Earth (1 R$_{\oplus}$) and Neptune ($\sim$4 R$_{\oplus}$), it was unclear whether they represent large-scale terrestrial planets (super-Earths) or small versions of Neptune (mini-Neptunes). Characterizing these planets is in particular challenging because we do not have similar objects in the Solar System, and they lie in a mass-regime where uncertainties in the Equation of State (EOS) and the material's distribution are the largest \citep[e.g.][]{Baraffe2008,Vazan2016}. 
\par

Determining the exact planetary structure and composition is challenging due to the intrinsic degeneracy, i.e., exoplanets with very different interiors can have identical masses and radii \citep[e.g.][]{Rogers2010,Lopez2014,Dorn2015,Dorn2017}. 
Despite this inherent degeneracy, the least dense possible interiors for a given bulk composition can be derived. These represent end-member interiors that can be compared to observed exoplanets. For example, the lowest density among all rocky (silicate) interiors is the one of MgSiO$_3$. 
Based on the density of the idealized composition of MgSiO$_3$,  
previous studies suggest that most of the planets with radii larger than 1.6 R$_\oplus$ have too low densities to be consistent with purely rocky interiors \citep{Rogers2015, Weiss2014}, and therefore they are expected to contain volatiles. More specifically, \citet{Rogers2015} employed a hierarchical Bayesian statistical approach to determine threshold radii of various rocky compositions. The threshold radius of a given composition represents the radius above which a planet has very low probability to be of that specific composition. \citet{Rogers2015} used a sample of 22 short period (up to 50 days) \textit{Kepler} planets with Radial Velocity follow-ups. For purely rocky exoplanets, a threshold radius of 1.6 R$_\oplus$ was found. 

Interestingly, the distribution of observed radii of small exoplanets suggests a bimodal shape of planetary sizes \citep{Fulton2017, Fulton2018}. A gap found at radii 1.5 - 2.0 R$_\oplus$ splits the population of close-in planets (orbital period shorter than 100 days) into two regimes: planets with $R_p< 1.5$ R$_\oplus$ and planets with $R_p=2.0-3.0$ R$_\oplus$. This paucity in the distribution might be explained by photo-evaporation of their volatile atmospheres \citep{Owen2017,VanEylen2017, Lopez2014}. 

Generally, in volatile-rich planets, the thickness of the gaseous envelope depends on the mass fraction of the light elements, the envelope's metallicity, and the temperature profile of the planet. 
These parameters and the characteristics of the underlying deeper layers determine the planet's density. Similarly to pure-rocky planets, end-member interior models for volatile-rich compositions exist. For example, a planet with a mass fraction of 2\% of H-He is expected to have the lowest density  when the envelope's metallicity is low, and the temperatures are high. 
In this paper, we build on the statistical methodology of \citet{Rogers2015}, and determine different threshold radii for small and intermediate-size planets, accounting for the possibility of gaseous envelopes with different metallicities and internal structures. 


\section{Methods}
\label{sec:Methods}
%

\subsection{Exoplanet Data}
\label{subsec:Data}

\begin{figure}

	\includegraphics[width=\columnwidth, trim={0.0cm 0.0cm 0.0cm 0.0cm},clip]{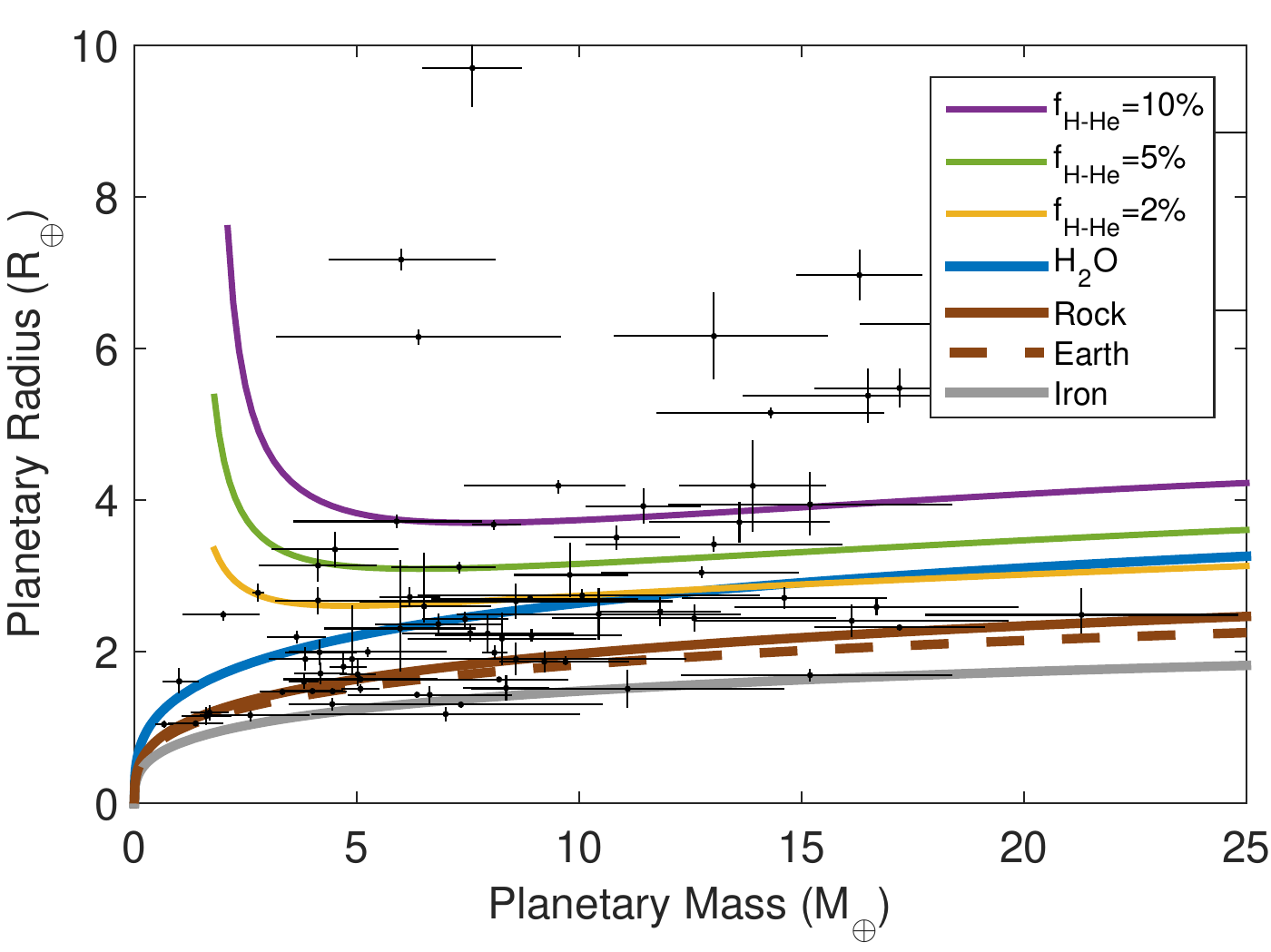}	
    \caption{The planet M-R diagram. The black dots with the error-bars are the planets we used in the analysis. The colored curves are examples of M-R relations for various theoretical compositions (see section \ref{subsec:Mass-Radius Relation} for details). The three M-R theoretical curves for planets with H-He atmospheres correspond to a semi-major axis of 0.1 AU, and a homogeneously mixed planetary envelope ({\it scenario-1}) with Z=0.2, as defined in Equation \ref{eq:Z}. }
    \label{fig:Smpl}
\end{figure}

To date (August 2018) there are more than 3700 confirmed exoplanets. Using the \textsl{exoplanet.eu} database, we select transiting planets with \textit{RV} or \textsl{TTV} follow-ups, with radii up to 10 R$_\oplus$ and masses up to 25 M$_\oplus$.
Planets with large uncertainties in the measured mass/radius (larger than 50\%), as well as planets with a measured uncertainty larger than 1 R$_\oplus$ and/or 4 M$_\oplus$ are excluded. That leaves us with a sample of 83 planets.
Figure \ref{fig:Smpl} shows the planetary sample, and the theoretical M-R relation curves (see section \ref{subsec:Mass-Radius Relation} for details). 
The corrections for data completeness are not considered since this study does not rely on the absolute frequency of planets, instead it is the measurement uncertainties that are relevant.


\subsection{The Mass-Radius Relation}
\label{subsec:Mass-Radius Relation}

Several theoretical M-R relations for various compositions have been derived by several groups \citep[e.g.][]{Seager2007,Zeng2013,Marcus2010,Lopez2014}. 
We use various theoretical compositions, such as pure H$_2$O, pure MgSiO$_3$, Earth-like composition (32\% Fe, 68\% silicate), and pure Fe, based on \citet{Seager2007}. 

In addition to these compositions, we construct a series of planetary models with rocky cores and volatile envelopes. The volatiles assumed in the envelope include hydrogen, helium, and water. 
For simplicity, we consider two end-member scenarios for the planetary structure. These two structures bracket the expected radii for a given planetary composition: 

\begin{enumerate}
	\item In {\it scenario-1} a rocky core is surrounded by an envelope consisting of H-He and water.  The hydrogen, helium and water are assumed to be homogeneously mixed.  
	\item In {\it scenario-2} we assume a completely differentiated structure in which the rocky core is surrounded by an inner pure H$_2$O layer and an outer layer composed of pure H-He. For that case the envelope corresponds to the 2-layer structure of water and H-He. 
\end{enumerate}

In both scenarios, we use the SCVH EOS \citep{Saumon1995} for H-He, with the H-He ratio being 72\% H to 28\% He. For the rocky core we use the EOS of MgSiO$_3$ \citep{Seager2007}. In {\it scenario-1} the water EOS is based on ANEOS by \citet{Thompson1990} (see \citet{Venturini2016} for details) while in {\it scenario-2} the water EOS is based on QEOS presented by \citet{More1988} \citep[see][for details]{Vazan2013}. 
The reason for using different water EOS is linked to the fact that we use two different codes for the two structures. 
This however, does not impact the inferred M-R relation, as presented in Figure \ref{fig:H2OEOS}. %
The Figure shows the M-R relation of a pure-water planet using ANEOS with a surface temperature T=500 K, QEOS with T=300 K and QEOS with T=1200 K alongside with \citet{Seager2007} polytropic EOS and \citet{Wagner2011} EOS. 
The three upper curves, corresponding to QEOS at 300 K and 1200 K and ANEOS at 500 K, are very similar (less than 1\% difference), suggesting that the inferred M-R relations should not be affected from using different water EOSs (see Section \ref{subsec:H2OEOS} for discussion). 
The M-R relations accounting for different mass fractions of H-He are derived by solving the standard internal structure equations, for the atmosphere we use the irradiation model of \citet{Guillot2010}. More details on the structure model can be found in Appendix \ref{AtmMdl} and references therein.  

\subsubsection{Key parameters} 
\label{subsubsec:Key}

In both scenarios, the planetary composition is defined by two parameters: the mass fraction of H-He ($\fHHe$), and the mass fraction of water in the envelope ($Z$).
These mass fractions are given by: 
\begin{equation}
    \fHHe =\frac{M_{\text{H-He}} }{M_{\text{H-He}} +M_{\text{H$_2$O}} + \Mrock },  
		\label{eq:f}
\end{equation}

\begin{equation}
	Z=\frac{M_{\text{H$_2$O}} }{ M_{\text{H$_2$O}}+M_{\text{H-He}}},
		\label{eq:Z}
\end{equation} 
where $M_{\text{H-He}}$, $M_{\text{H$_2$O}} $, $\Mrock$ are the masses of H-He, water, and rock, respectively. 

The envelope's mass fraction (homogeneously mixed/differentiated) is then given by: 
\begin{equation}
	f_{\text{env}} = \frac{M_{\text{H-He}} +M_{\text{H$_2$O}} }{M_{\text{H-He}} +M_{\text{H$_2$O}} + \Mrock }.
		\label{eq:fatm}
\end{equation}
Similarly, we can define the mass fraction of water by:
\begin{equation}
	\fwater = \frac{ M_{\text{H$_2$O}} }{M_{\text{H-He}} +M_{\text{H$_2$O}} + \Mrock }, 
		\label{eq:fwater}
\end{equation}
and the rock mass fraction as: 
\begin{equation}
	\frock = \frac{ \Mrock }{\MHHe +\Mwater + \Mrock }.
		\label{eq:frock}
\end{equation}
Since $\fHHe + \fwater + \frock = 1$, the following relations can be derived:
\begin{equation} \label{waterZ}
\fwater = \bigg(\frac{Z}{1-Z} \bigg) \fHHe
\end{equation}

\begin{equation} \label{rockZ}
\frock = 1-\frac{\fHHe}{1-Z} 
\end{equation}

\begin{align} \label{water-rock}
f_{\text{W/R}} =& \, \frac{\fwater}{\frock}  \nonumber \\
=& \, \frac{Z  \, \, \fHHe}{1-Z-\fHHe}
\end{align}
where the latter stands for the planetary \textit{water-to-rock} mass ratio. 

It should be noted that in principle, one can choose different key parameters to define the planetary composition, such as the water-to-rock ratio. 
Other possible definitions for a planetary structure model with alternative key parameters but the same composition (rock, water, H-He) are not physically different from our models, but differ by the way the mass fractions are defined. 

Since we focus on intermediate- and low- mass planets, we consider $\fHHe$ values between 2\% and 10\% and $Z$ values between 0 and 0.7. Such H-He mass fractions are the minimum expected for mini-Neptunes \citep[e.g.,][]{Venturini2017} and 
the envelope's metallicities expected for Neptune-like planets \citep[e.g.,][]{Helled2011}. 
The mean density of planets with larger $\fHHe$ is significantly lower than the typical density of the observed exoplanets with masses up to 25 M$_\oplus$. As a result, we do not consider $\fHHe$ larger than 10\%.

\subsubsection{M-R Relation of volatile-rich planets}
\label{subsubsec:fZ}

The range of possible M-R relations for planets with volatile envelopes is presented in Figure \ref{fig:VarZf}. 
The row and the color corresponds to $\fHHe$ (2\%, 5\%, 10\%), and the column corresponds to various $Z$ (0.1, 0.2, 0.4, 0.7). Each subplot shows the range of possible M-R relations for a given $\fHHe$ and $Z$.  
The upper limit is determined by the inferred M-R relation of a fully differentiated structure (\textit{scenario-2}), while the lower one is given by the fully mixed models (\textit{scenario-1}). 
\par

It is found that the range of possible models increases with increasing $\fHHe$ and $Z$ values, but decreases with increasing planetary mass as presented in Figure \ref{fig:VarZfsum1} (various colors present different $\fHHe$ and line style represents various $Z$). 
Figure 4 shows the absolute difference  between \textit{scenario-2} and \textit{scenario-1} as a function of planetary mass. We confirm that the distribution of elements within the interior of intermediate mass planets has a large effect on the radius  \citep[e.g.][]{Baraffe2008,Vazan2016}. This should be accounted for when characterising planets in this mass/size regime. 

As can be seen in the figures, $\fwater$ increases with $Z$ (see equations \ref{eq:Z} and \ref{eq:fwater}), as well as with increased $\fHHe$. 
While this might be unintuitive, this behaviour is a result of  our composition definition: increasing $\fHHe$ while keeping a constant $Z$ leads to an increase in $\fwater$ (Eq. \ref{waterZ}), and therefore to a decrease in $\frock$ (Eq. \ref{rockZ}). 

The planetary temperature must be included when studying the M-R relation of planets consisting of volatile materials \citep[e.g.,][]{Lopez2014,Swift2012}. The radii of planets with H-He atmospheres are larger for higher temperatures. 
It should be noted, however, that most of the planets in the sample have equilibrium temperatures of $\sim$500 K and semi-major axes of $\sim$0.1 AU. Therefore, we use a semi-major axis of 0.1 AU as the default when deriving the M-R relations. 
A more delicate analysis where the temperatures are derived for each planet individually, accounting for semi-major axes and different albedos was also performed and is presented in Section \ref{sec:AdjT}. 

\subsection{The Statistical Analysis}
\label{subsec:Statistical Analysis}

We aim to determine the probability of a given planet to be denser than a given composition, based on the M-R relations. 
The theoretical M-R relations are used to define the transition  between different possible compositional regimes. For example, the solid-brown curve in Figure \ref{fig:Smpl} corresponds to MgSiO$_3$ (the least-dense silicate composition).
If a given planet is above the curve, it indicates that the planet is less dense than pure rock, and thus has some volatiles (e.g., water and/or H-He). 
The gray line corresponds to pure iron and is used as the highest density possible for terrestrial planets. Planets between iron and rock lines are likely to consist of a mixture of silicates and iron and can therefore be referred as "potentially rocky" \citep{Rogers2015}. 
A similar reasoning is applied for planets with H-He atmospheres: for example, a planet above our $\fHHe$ = 5\% curve is likely to have an atmosphere with $\fHHe >$ 5\%.

\subsubsection{Probability calculation}
\label{subsubsec:p-calc}

The measured values of the mass and radius and their uncertainties play a key role in the analysis; the larger the uncertainties, the larger the range of possible compositions. Constraining the planetary composition is performed as follows: if a given planet is located below a given M-R curve, it suggests that the planet is denser than a specific composition (i.e., consists of heavier elements). 
Due to measurement uncertainties, we define the probability \textit{p} of a planet to be below a given M-R curve. Planets with \textit{p} of $\sim$ 1 are very likely to be composed of elements that are denser than a particular structure, while planets with \textit{p} of $\sim$ 0 are likely to be composed of lighter (i.e., more volatile) materials. 

The measured values of mass and radius are assumed to have an asymmetric normal-like distribution: 
\begin{equation}
M_{pl} \sim \mathcal{N}(M,M_{err\pm}) = \begin{cases} \mathcal{N}(M,M_{err-}) & M_{pl}<M \\ \mathcal{N}(M,M_{err+}) & M_{pl}>M \end{cases},
\label{eq:Merrs}
\end{equation}
\begin{equation}
R_{pl} \sim \mathcal{N}(R,R_{err\pm}) = \begin{cases} \mathcal{N}(R,R_{err-}) & R_{pl}<R \\ \mathcal{N}(R,R_{err+}) & R_{pl}>R \end{cases},
\label{eq:Rerrs}
\end{equation}
where $M,R$ are the measured radius and mass, respectively, and $M_{err\pm}, R_{err\pm}$ are the corresponding measurement uncertainties. 

For each measured mass-radius pair, we randomly sample 10,000 physically plausible synthetic planets. 
The simulated values are asymmetrically normally distributed using the measured data, $M_p\sim \mathcal{N}(M, M_{err\pm})$, $R_p\sim \mathcal{N}(R, R_{err\pm})$, as defined in equations \ref{eq:Merrs}-\ref{eq:Rerrs}. 

This simulated sample is used to determine \textit{p}, the probability of a given planet to be in a desired M-R region. 
The probability of a planet to be denser than a given composition is given by the fraction of simulated points that found below a given M-R curve:

\begin{equation}
	p=\frac{ \#\textrm{points below the curve} }{\#\textrm{simulated points}} .
		\label{eq:p}
\end{equation}

Simulated points that fall below the iron curve (gray line in Figure \ref{fig:Smpl}) and points with negative radius and/or mass are unphysical, and are therefore excluded. 
We infer the probability \textit{p} for each individual planet in our planet sample. This procedure is then repeated for every M-R curve separately. In order to ensure that the inferred value of $p$ does not depend on the size of the simulated sample, we have run cases with smaller samples (1,000 and 500) and got similar $p$ values, suggesting that our derived value is robust. 

\subsection{Threshold Radius R$_{th}$}
\label{subsubsec:Threshold}

We investigate whether there is a sharp threshold on the distribution of $p$ in terms of the planetary radius for different assumed compositions, which should result in a step-function. 
Similarly to \citet{Rogers2015}, we represent the distribution of $p$, as a function of $R$ with a step-function (upper panels in Figures \ref{fig:Trip1}-\ref{fig:Trip2}):
\begin{equation}
    \Theta(R_p,R_{th})=\begin{cases} 1, & R_p < R_{th}, \\ 0, & R_p \ge R_{th}, \end{cases} ,
	\label{eq:step}
\end{equation}
where $R_p$ is the planetary radius and $R_{th}$ is the threshold radius, to be found. 
Slightly more complex functions with a gradual transition has been explored by \citet{Rogers2015}, and it was found that the simple step-function essentially coincides with the best fit of a gradual transition.
As discussed above, R$_{th}$ represents the lower limit of a planetary radius allowing the planet to be denser than a given composition. 


For every composition, we derive an ensemble of different possible threshold radii r$_{th}$, in order to find the best fit for $R_{th}$. 
We search for a radius R$_{th}$ that minimizes the mean squared error ($MSE$) between the $p$ points and the $\Theta(R_p,r_{th})$ curve. Formally, we minimize the term: 
\begin{equation}
	MSE(r_{th})=\frac{1}{j} \sum_j \left|\Theta(R_j,r_{th}) - p_j\right|^2  ,
		\label{eq:MSE}
\end{equation}
where \textit{j} runs over a fixed number of planets (83), $R_j$ are the measured planetary radii, and $p_j$ are the corresponding probabilities to be below a certain M-R curve (see upper panels of Figures \ref{fig:Trip1}-\ref{fig:Trip2}). 
The quality of the fit is inversely proportional to the value of $MSE$, and therefore the threshold radius is $r_{th}=R_{th}$ that minimizes the function $MSE (r_{th})$. Mean squared error ($MSE$) is shown in the lower panels of Figures \ref{fig:Trip1}-\ref{fig:Trip2}. 
The location of $R_{th}$, corresponding to the minimal value of $MSE$, is represented by the red dashed line. 
The value of $MSE$ starts to drop in a radius region where $p$ transits form $\sim 1$ to $\sim 0$ (Figure \ref{fig:Trip2}). 
The quality of the fit to $R_{th}$ is better for lower values of $MSE$.
The spread of planets in a transition region with $0<p<1$ correlates with the width of the trough, and therefore with the uncertainty on estimating $R_{th}$ as we discuss below. 


The uncertainties on the threshold radii were derived using a "Bootstrap method". 
In this method we use random sampling with replacement, based on pairs of the measured $R_p$ and the corresponding calculated $p$. 
As discussed before, $R_{th}$ is the value that minimizes the $MSE$ function. In the bootstrapping method, we recalculate $MSE$ and then $R_{th}$, using sub-samples of the original ($p,R_p$) set. 
In the procedure, which is repeated 10,000 times, we are sampling with a replacement 83 pairs ($p_i,R_{p,i}$) from the original ($p,R_p$) sample. The simulated sub-sample ($p_i,R_{p,i}$) has the same length as the original one, but includes repetitions of a random number of values. 
For each sub-sample ($p_i,R_{p,i}$) we calculate the $MSE_i$, and find a  corresponding threshold radius $R_{th,i}$. This procedure creates 10,000 $R_{th,i}$ values of the sample statistics. 

The sample statistics may include some unphysical extreme values. Therefore, we exclude the lowest 2.5 percent and the highest 2.5 percent of the simulated set for $R_{th,i}$. 
Then, we construct a Cumulative Distribution Function ($CDF$) from the simulated $R_{th,i}$ values (see Figure \ref{fig:cdf}). 
The 50-th percentile of the $CDF$ corresponds to the mean value of the $R_{th}$ (shown by the dark gray dashed line in Figure \ref{fig:cdf}), while the 16-th and 84-th percentile values (shown in the light gray dashed lines) correspond to the standard error, and therefore to the formal uncertainties on $R_{th}$. Note that the asymmetric form of the $MSE$ in Figure \ref{fig:Trip2} leads to asymmetric uncertainties ($R_{err+}$ and $R_{err-}$ on Figure \ref{fig:cdf}).  
In a few cases the boundaries are inconclusive due to lack of separation between the 16-th or/and 84-th percentile and the 50-th percentile. For theses cases the method cannot provide a meaningful uncertainty (being referred as $n/a$). 

\section{Results}
\label{sec:Res}

The calculated threshold radii and their uncertainties for different possible compositions are summarized in Tables \ref{tab:Rocky}-\ref{tab:SumStab}. 
The results for the threshold radii $R_{th}$ are presented in Figures \ref{fig:Trip1}-\ref{fig:Trip2}, where the top panels show the probability to be denser than a given composition \textit{p} as a function of planetary radius $R_p$ for different theoretical compositions.  

\subsection{Planets without volatiles}
\label{subsec:RockRes}
First, we confirm the result of \citet{Rogers2015} that planets with radii larger than 1.6 R$_\oplus$ are not pure-rock. 
We find that planets above 1.4 R$_\oplus$ cannot have Earth-like compositions (32\% iron, 68\% silicate) and therefore have to contain larger fraction of silicates or lighter materials. 
The threshold radii for volatile-poor planets are summarized in Table \ref{tab:Rocky}. 

\subsection{Planets with volatile envelopes}
\label{subsec:GasRes}

We find that most of the planets larger than 2.6 R$_\oplus$ cannot be pure-water worlds. As expected, the planetary radius is typically increasing with $\fHHe$. 
The results for planetary models with gaseous envelopes are listed in Table \ref{tab:VarZvarF}. 
For planets with homogenous envelopes (\textit{scenario-1}), we find that $R_{th}$ is typically lower than $R_{th}$ of the differentiated structure (\textit{scenario-2}) for the same bulk composition (see Section \ref{subsec:Mass-Radius Relation} for details). 
That is illustrated in Figure \ref{fig:VarZf}, which shows the boundaries derived by the two scenarios. 

It is interesting to note that while in the fully mixed case ({\it scenario-1}) $R_{th}$ decreases with $Z$, the opposite occurs for the differentiated structure ({\it scenario-2}). 
This is a consequence of the way we built our planets: in {\it scenario-1} increasing the envelope's metallicity leads to a contraction of the envelope due to self-gravity, and thus to a smaller radius. 
In {\it scenario-2}, the volume of pure H-He is constant for a given $\fHHe$, and the only effect of increasing $Z$ is an increase in the water-to-rock ratio (Eq. \ref{water-rock}). 
Since the amount of water, which has a lower density than rocks, is increased at the expense of reducing the abundance of rocks in {\it scenario-2} the planet's radius increases with increasing $Z$. 
\par

For the various compositions and internal structure we consider (Table 2), we find that the threshold radii in the range between 2.5--4.3 R$_\oplus$ depending on the chosen scenario, $\fHHe$ and $Z$. 
We find that a planet above 2.6 R$_\oplus$ must have significant amounts of H-He and therefore can be classified as mini-Neptunes. 
There is degeneracy between the threshold for pure-water worlds ($R_{th}=2.58^{+0.05}_{-0.05}$ R$_\oplus$) and mini-Neptuns with high ($Z=0.7$) atmospheric metalicity ($R_{th}=2.52^{+0.03}_{-0.14}$ R$_\oplus$). 
Although exact composition of planets with these sizes cannot be inferred exactly, 
we can conclude that planets with radii larger than $\sim$ 2.6 R$_\oplus$ are likely to consist of H-He atmospheres. In addition, we find that planets with sizes $R_p \gtrsim$ 4 R$_\oplus$ are likely to have significant H-He atmospheres (more than 10\% of the planetary mass). 
It should be noted that the planets in our sample are significantly hotter than the Solar System's Uranus and Neptune (see Figure \ref{fig:AlbHist}). 
Therefore the planets that are found to have H-He envelopes are hot- and warm- Neptunes/mini-Neptunes, i.e., with compositions similar to that of Uranus/Neptune but with    
 a larger radius due to stellar irradiation \citep{Baraffe2006}. 

\section{Sensitivity of the results to the model assumptions} 

\subsection{Sensitivity to the EOS of water} 
\label{subsec:H2OEOS}

\begin{figure}
		\includegraphics[width=\columnwidth, trim={3.0cm 9.0cm 3.0cm 9.5cm},clip]{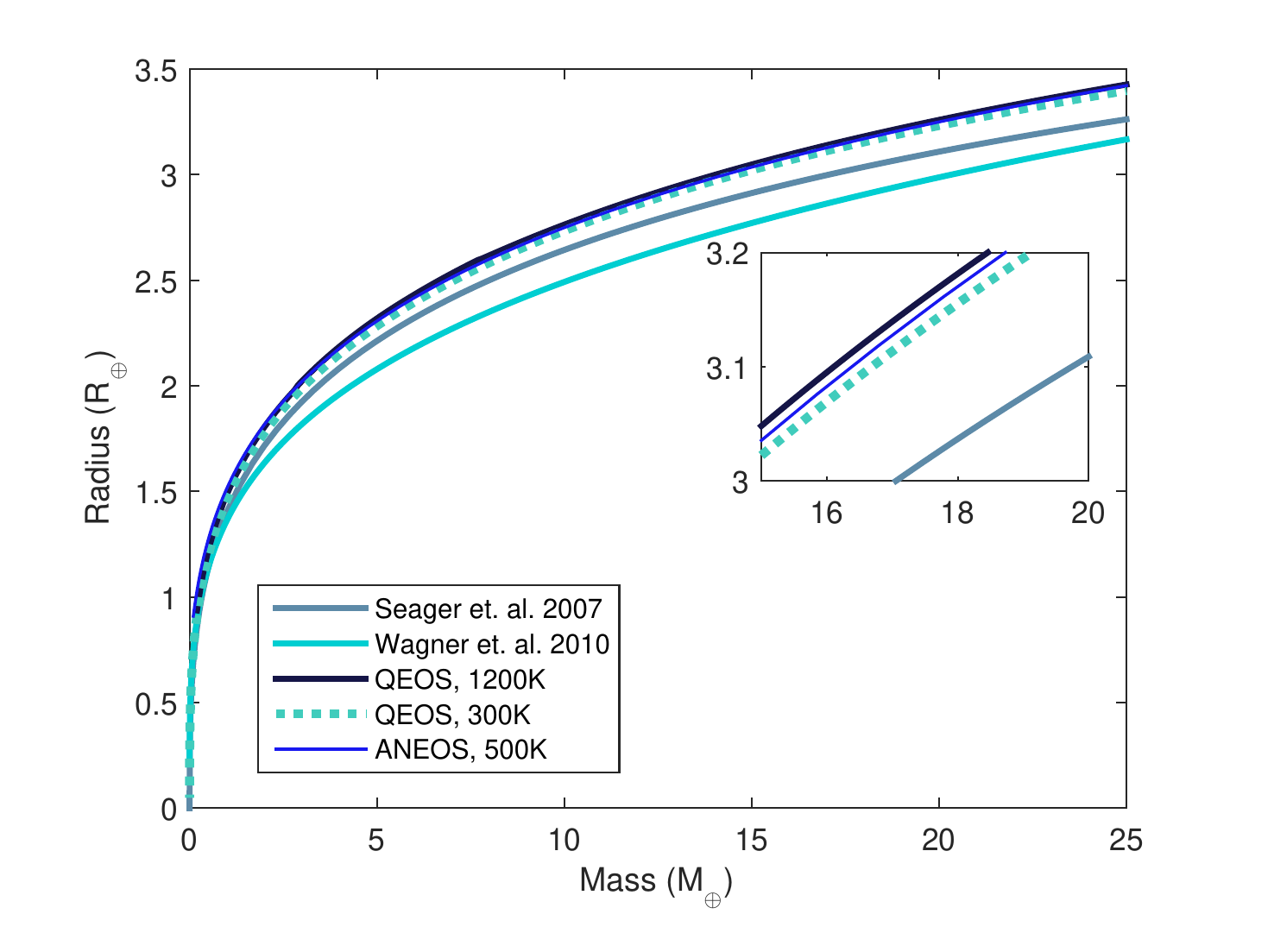}
    \caption{M-R relations for pure water planets, using different EOSs for water. In this study we use the polytropic EOS of \citet{Seager2007} for a pure-water planet. Other EOSs presented here are \citet{Wagner2011} EOS, QEOS assuming surface temperatures of T=300K and T=1200 K \citep{More1988}, and ANEOS \citep{Thompson1990} with a surface temperature of $T=500 K$. In this work, ANEOS was used in \textit{scenario-1} and QEOS was used in \textit{scenario-2} (see text for details).}
    \label{fig:H2OEOS}
\end{figure}

\begin{figure}
	\includegraphics[width=\columnwidth,trim={0.0cm 6.0cm 0.0cm 5.5cm},clip]{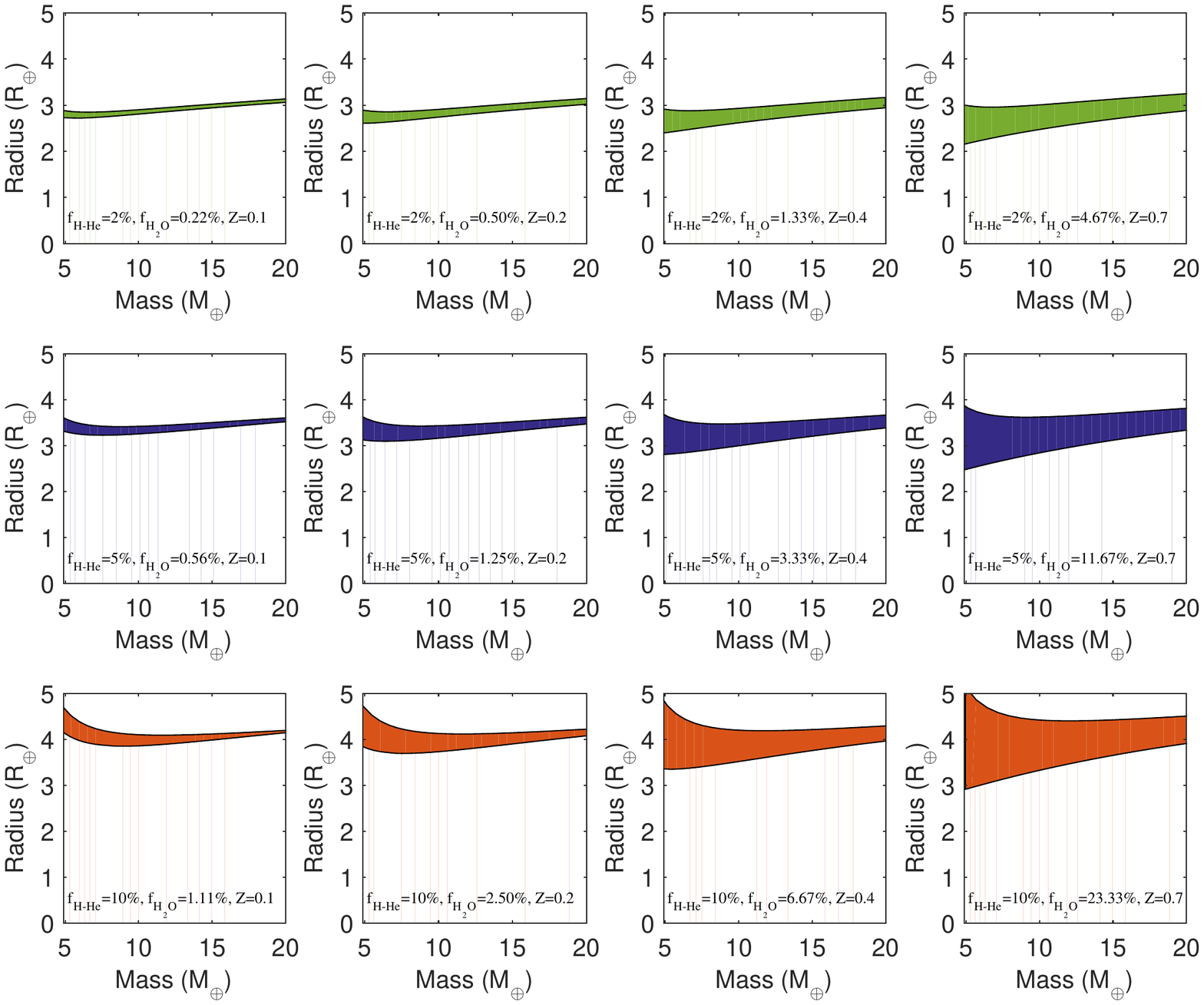}	
    \caption{M-R relation ranges for structure models with a rocky core and various fractions of hydrogen and helium ($\fHHe$) and atmospheric metalicities ($Z$) (see text for details). The lower limit on the radius corresponds to models with the water being mixed in the H-He envelope ({\it scenario-1}), while the upper one corresponds to the fully differentiated structure ({\it scenario-2}). 
    The mass fractions of H-He $\fHHe$ and water $f_{H_2O}$, and the assumed envelope's metallicity ($Z$) are indicated in each panel.}
    \label{fig:VarZf}
\end{figure}

\begin{figure}
	\includegraphics[width=\columnwidth,trim={3.95cm 9.0cm 4.0cm 6.5cm},clip]{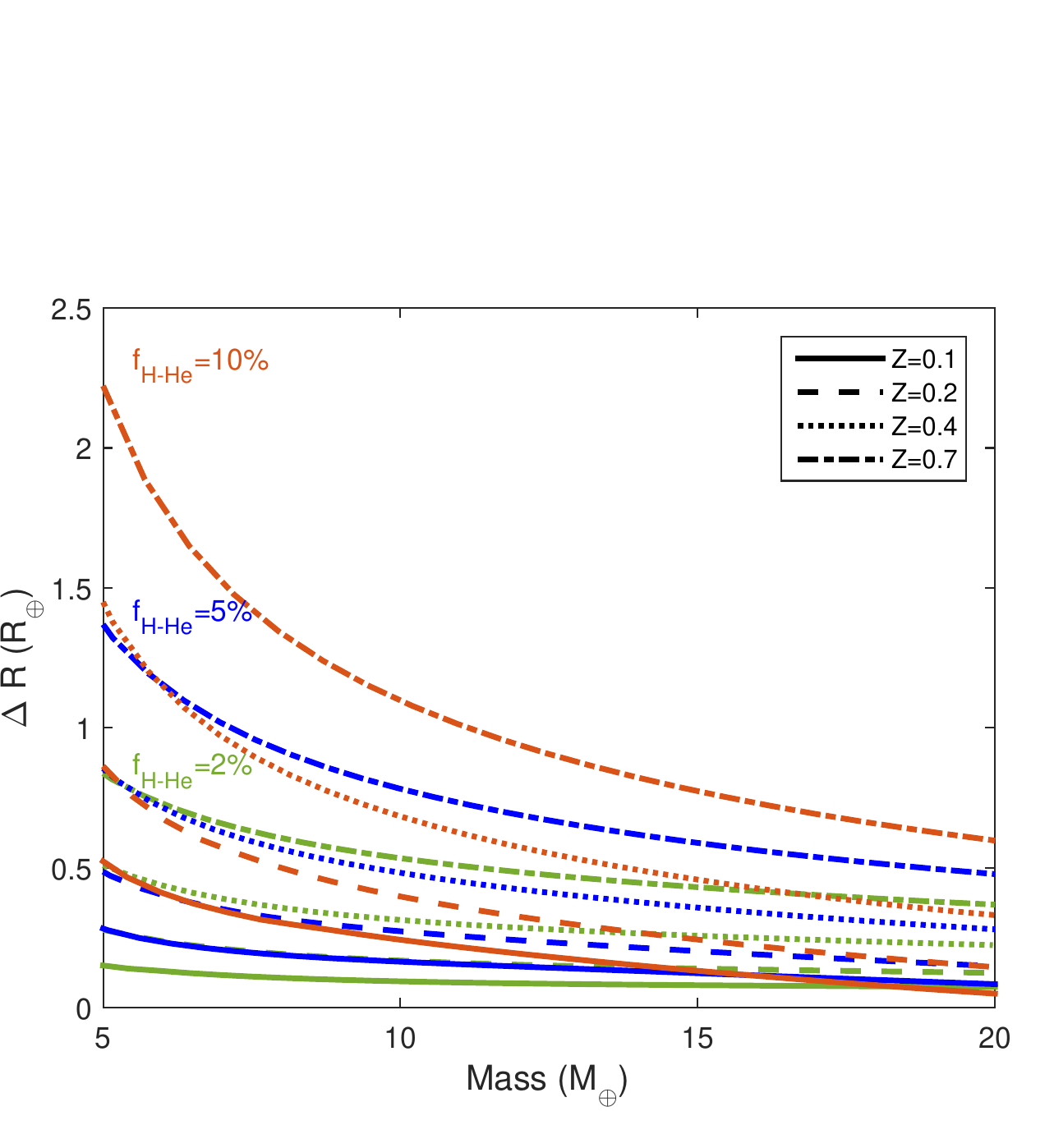}	
    \caption{The differences in the inferred threshold radius between \textit{scenario-1} and \textit{scenario-2}. The colors represent percentage of H-He ($\fHHe$), and the line style the assumed atmospheric metallicity $Z$.}
    \label{fig:VarZfsum1}
\end{figure}

\begin{figure*}
	\includegraphics[width=\textwidth, trim={2.5cm 4.0cm 2.5cm 0.3cm},clip]{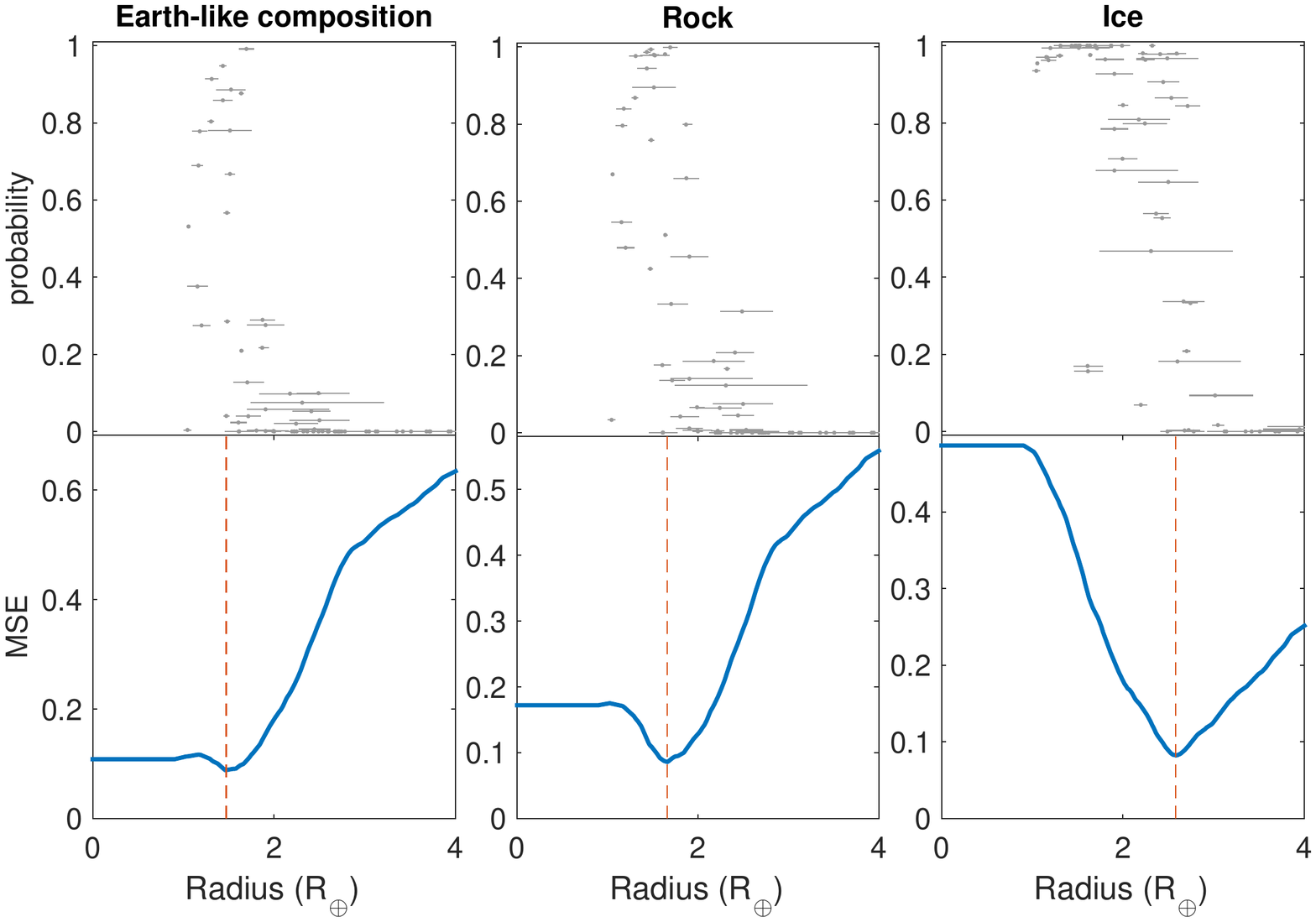}
    \caption{The probability for a planet to be denser than a given composition, and the corresponding threshold radius R$_{th}$. {\bf Top:} the probability to be denser than a given theoretical composition (Earth-like, pure-rock, pure-water) as a function of planetary radius.  {\bf Bottom:} Mean Squared Error (MSE) of the fitted threshold step-function. The red dashed line corresponds to the best fit (see subsection \ref{subsubsec:Threshold}). }
    \label{fig:Trip1}
\end{figure*}

\begin{figure*}
	\includegraphics[width=\textwidth, trim={2.5cm 4.0cm 2.5cm 0.3cm},clip]{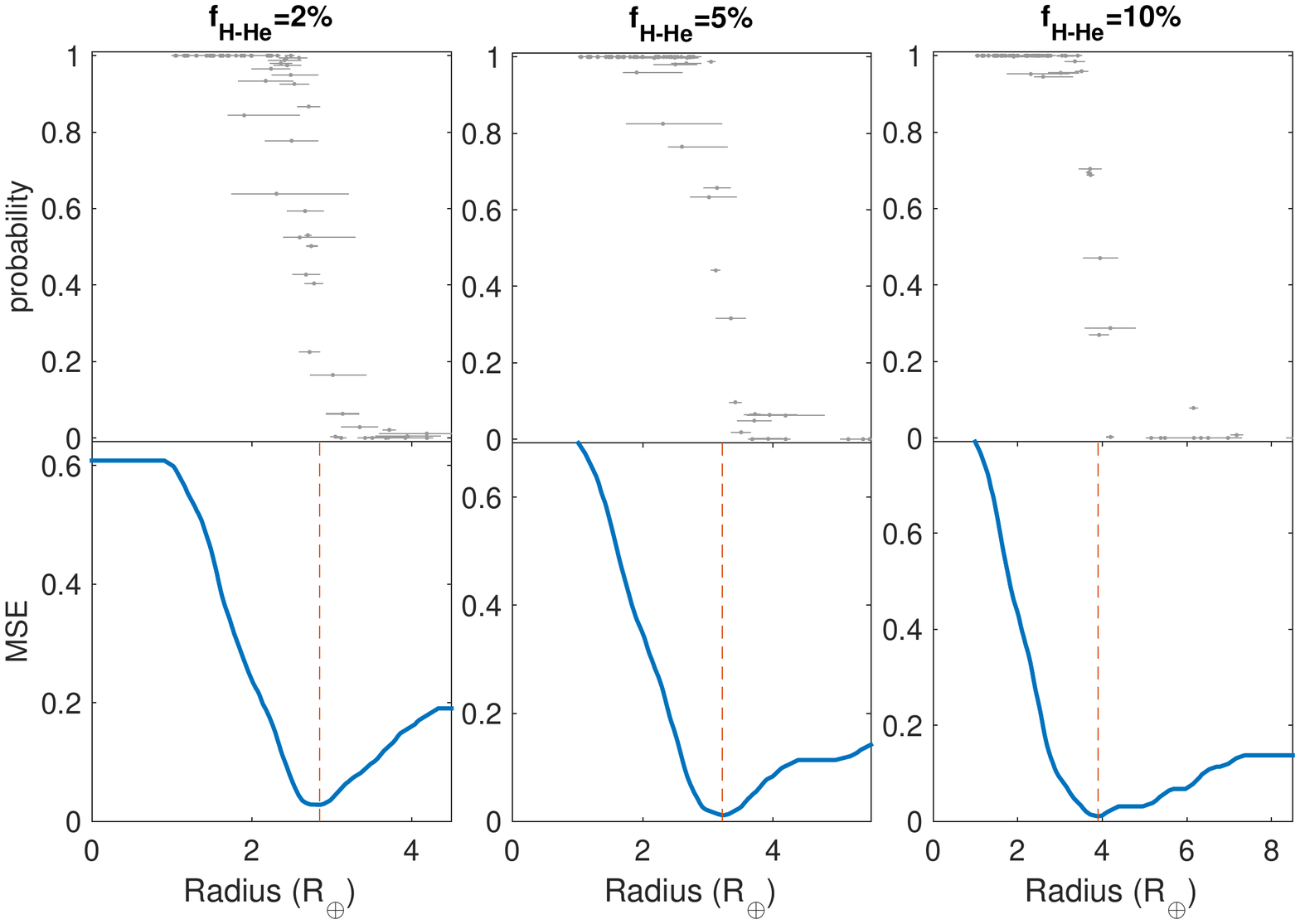}
    \caption{Same as Figure \ref{fig:Trip1}, for models with a rocky core surrounded by an envelope consists of homogeneously mixed H-He and water ({\it scenario-1}) with $Z=0.2$. The percentage in the title is the mass fraction of H-He ($\fHHe$). }
    \label{fig:Trip2}
\end{figure*}

\begin{figure}
	\includegraphics[width=\columnwidth,trim={3.5cm 9.0cm 4.0cm 7.0cm},clip]{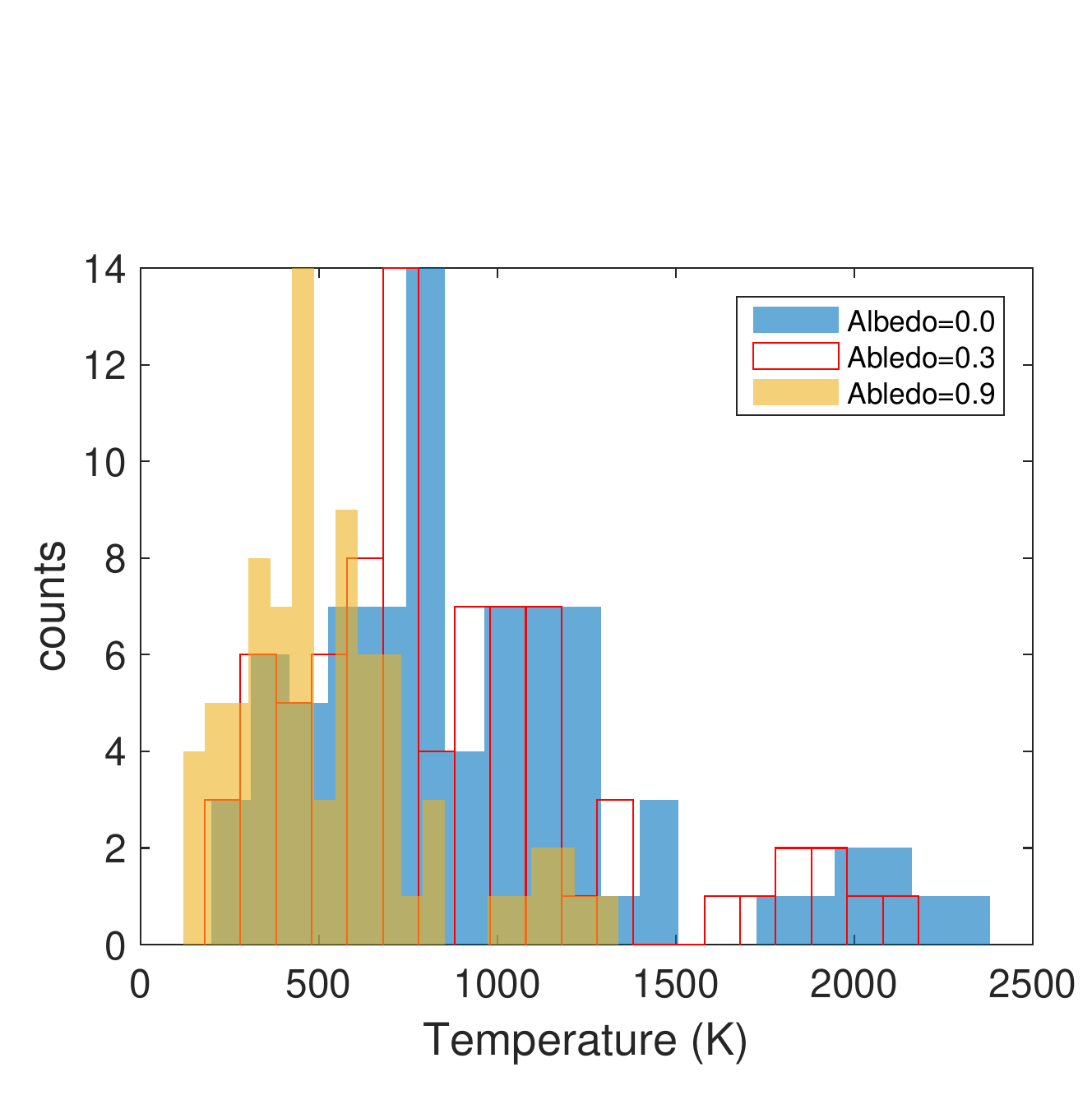}	
    \caption{Planetary equilibrium temperature histograms for three different albedos: A=0.0, A=0.3, and A=0.9. }
    \label{fig:AlbHist}
\end{figure}

The EOS of water (as well as other elements) is still not perfectly known, especially in the high pressure-temperature regime. 
In the case of pure-water planet, the default EOS we use for water is that of \citet{Seager2007}. However, differences in the water EOS could lead to differences in the derived inferred radii. 
As a result, we investigate the sensitivity of R$_{th}$ to the assumed water EOS.  A comparison of the M-R relation for pure-water planets using three different water EOSs is presented in Figure \ref{fig:H2OEOS}.  
Additional curves using the \citet{Wagner2011} and \citet{More1988} EOSs for water, and different  effective temperatures are also presented.  
The differences between the EOSs is mainly linked to the different assumed bulk modulus ($K_0$ and $K'_0$) of the planetary ices, but is in general relatively small. 
Therefore, the inferred threshold radii are relatively insensitive to the water EOS. The derived R$_{th}$ assuming various EOSs for water are presented in Table \ref{tab:H2OcompEOS}. 
We find that R$_{th}$ changes from 2.58$^{+0.05}_{-0.05}$ R$_\oplus$ to 2.63$^{+0.10}_{-0.05}$ R$_\oplus$ for the most extreme cases. 
As shown in Figure \ref{fig:H2OEOS}, the differences between ANEOS and QEOS for water are negligible. We can therefore conclude that the differences between the M-R relations derived in \textit{scenario-1} and \textit{scenario-2} are linked to the distribution of elements (i.e., the assumed internal structure) and are not affected by the choice of the water EOS. 


\subsection{Sensitivity to the equilibrium temperature and planetary albedo}
\label{sec:AdjT}

The M-R relation derived for compositions (and structures) with significant amount of volatiles depends on the planet's equilibrium temperature.  Information on the stellar and orbital properties of the system, such as stellar temperature, stellar radius and semi-major axis, can be used to calculate the planetary equilibrium temperature ${ T }_{ eq }$.  
The equilibrium temperature is given by: 
\begin{equation}
{ T }_{ eq }={ T }_{ \odot  }{ \left( 1-A \right)  }^{ 1/4 }\sqrt { \frac { { R }_{ \odot  } }{ 2D }  } ,
	\label{eq:Teq}
\end{equation}
where ${ T }_{ \odot }$ and ${ R }_{ \odot }$ are the stellar temperature and radius, respectively, $D$ is the semi-major axis, and $A$ is the planetary albedo. 
In this study we set the default case for a Sun-like star, semi-major axis of 0.1 AU, and albedo of 0 (i.e., black body, full absorption). Our default values correspond to ${ T }_{ eq }$= 770 K. 

We next investigate the effect of the semi-major axis on the threshold radius (assuming a constant albedo of $A=0$). 
We derive new M-R relations for {\it scenario-1} with 5\% of H-He and repeat the analysis for the new curves using various semi-major axes. 
We find that R$_{th}$ varies form 2.9 R$_\oplus$ to 3.5 R$_\oplus$ for semi-major axes between 0.05 AU and 0.5 AU, with the larger radius corresponding to the smaller radial distance. The inferred threshold radii for different assumed temperatures are summarized in the first three rows of Table \ref{tab:Summary3}. Since a semi-major axis of  $\sim$0.1 AU corresponds to the majority of the planets in the sample, we use this as the default value. 

The planetary albedo depends on many variables such cloud layers and chemical composition. To explore the sensitivity of the inferred R$_{th}$ to the assumed albedo we perform the analysis assuming  three different albedo values: $A=0$ (low, full absorption), $A=0.3$ (Earth-like, intermediate), and $A=0.9$ (high), while keeping a semi-major axis of 0.1 AU. A histogram of the derived planetary temperature for the different albedos is presented in Figure \ref{fig:AlbHist}. As expected, higher albedo leads to a lower equilibrium temperature. 

In a second test, instead of using a single M-R relation with a constant temperature for the entire planetary sample, as done in Section 3.2, we adjust the M-R relation for each planet individually using the calculated equilibrium temperature and assumed albedo. 
We then calculate $p$ for each planet using the individually calculated M-R curve. After the $p$ distribution for the planetary sample is derived, we find R$_{th}$ in the same fashion as described above.   
The results are summarized in the lower three entries of Table \ref{tab:Summary3}. We find that R$_{th}$ is relatively insensitive to the assumed albedo value. The cases with albedos of 0 and 0.3 are essentially identical (3.19 R$_\oplus$), 
and the inferred threshold radius is not very different even when using $A=0.9$ (2.97 R$_\oplus$). 

\subsection{Sensitivity to luminosity}
\label{subsec:Lum}
The planetary luminosity in our models is calculated using the luminosity fit of \citet{Rogers2010} 
(see Appendix for details).  
In this approach the luminosity varies as a function of planetary mass and radius in a range of $\sim 10^0-10^3 L_N$, where $L_N$ is Neptune's luminosity (377 GW). 
In order to test the sensitivity of the results to the assumed luminosity, we apply different constant luminosity values (between $\sim 10^{-1 }L_N$ and $10^3 L_N$) when constructing the planetary structure and their corresponding M-R relations. We apply the comparison to \textit{scenario-1} with $\fHHe$=5\% and Z=0.2. 
The comparison between the different cases is summarized in Table \ref{tab:varL}). As can be seem from the Table, R$_{th}$ changes only by up to 10\% for a range of luminosities that covers several orders of magnitude. We can therefore conclude that the inferred threshold radii are relatively insensitive to the assumed luminosity. 
\subsection{Sensitivity to the used data sample} 
\label{subsec:Stability test}

\begin{figure}
		\includegraphics[width=\columnwidth, trim={3.0cm 8.5cm 4.5cm 8.5cm},clip]{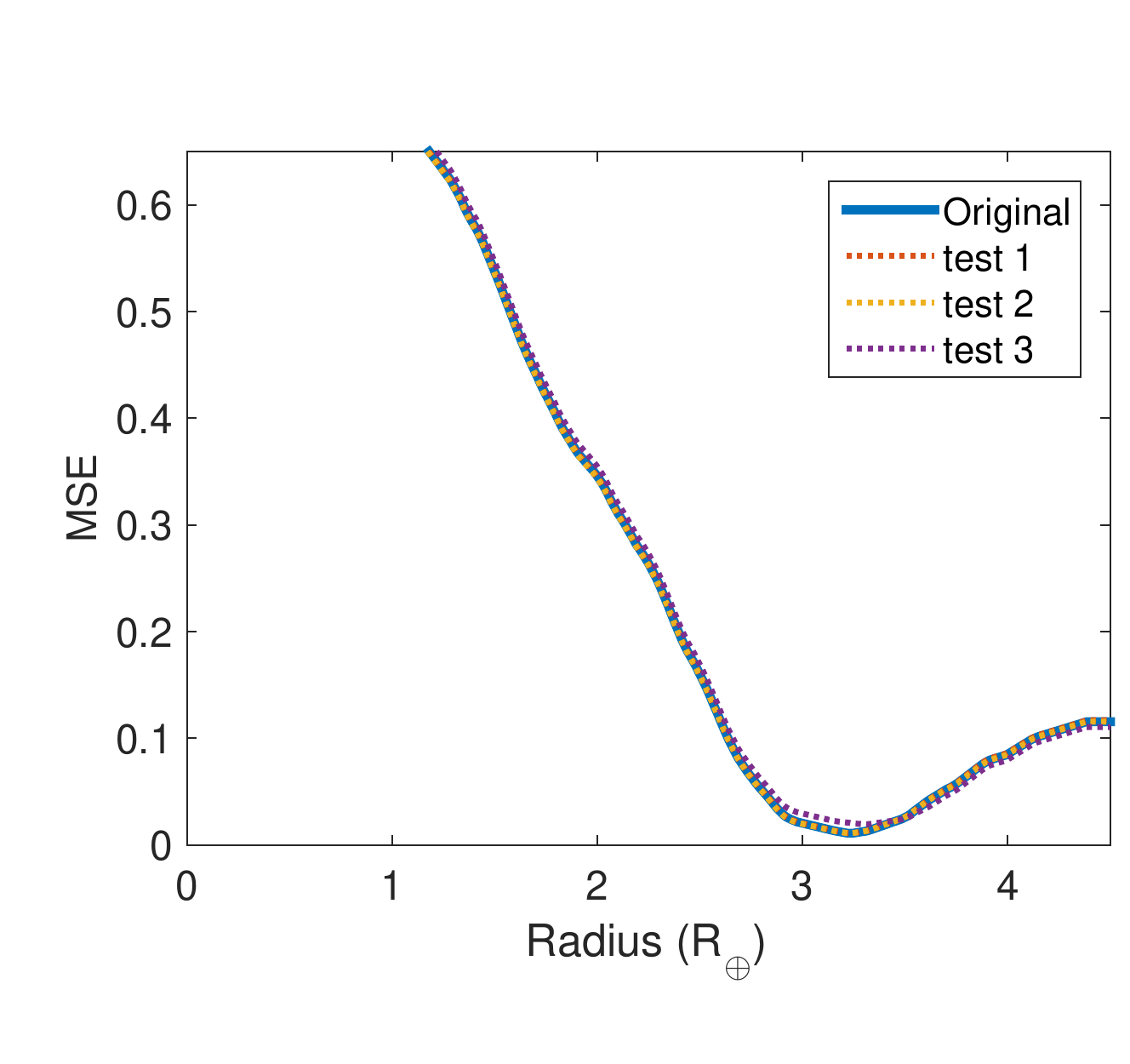}
    \caption{Sensitivity to the used data sample. Shown are the MSE for the original M-R data (solid line) versus the test (modified) cases (dotted lines). See text for details. }
    \label{fig:stabTest}
\end{figure}

Systematic observational biases can influence the determination of the planetary mass and/or radius. 
In order to investigate the sensitivity of our results to the used data, i.e., exact values of the measured mass and radius (and their uncertainties), we explore the dependence of R$_\oplus$ on the exact values of the masses and radii. We then randomly reassign the masses (with the corresponding mass uncertainties) 
for 20\% of the planets in the sample (16 planets in total), while keeping the same radii and their corresponding uncertainties. 
This way we ensure that the mass and radius distributions of the synthetic data sample are kept the same compared to the original data. 
We then repeat the analysis using the new partially randomized sample (modified data sample). This test has been performed three times in order to explore the robustness of the results. 
The results are presented in Figure \ref{fig:stabTest} and Table \ref{tab:SumStab}. We find that the threshold radius changes by less than 2\%, and we therefore conclude that our results are robust.  

\subsection{The existence of a threshold mass M$_{th}$?} 
\label{subsubsec:Mth}

We find that the existence of threshold radii is statistically significant, and in principle, one could expect to have a similar behaviour for the mass. 
Planet formation models predict that a heavy-element core starts to accrete H-He in significant amounts at around the so-called \textit{critical core mass}. 
The value of this mass depends primarily on the solid accretion rate, envelope composition, and opacity \citep[e.g.][]{Ikoma2000,Venturini2016, Venturini2017}, and represents the transition between gas-poor and gas-rich planets. 
At the moment, the estimates for this critical core mass range from less than 1 M$_{\oplus}$ to several Earth masses \citep[e.g.][]{Pollack1996,Brouwers2018}. 
Therefore, finding a threshold mass could provide important constraints on the physical conditions and dominating processes during planet formation. 

\begin{figure}
		\includegraphics[width=\columnwidth, trim={3.0cm 8.5cm 2.5cm 8.5cm},clip]{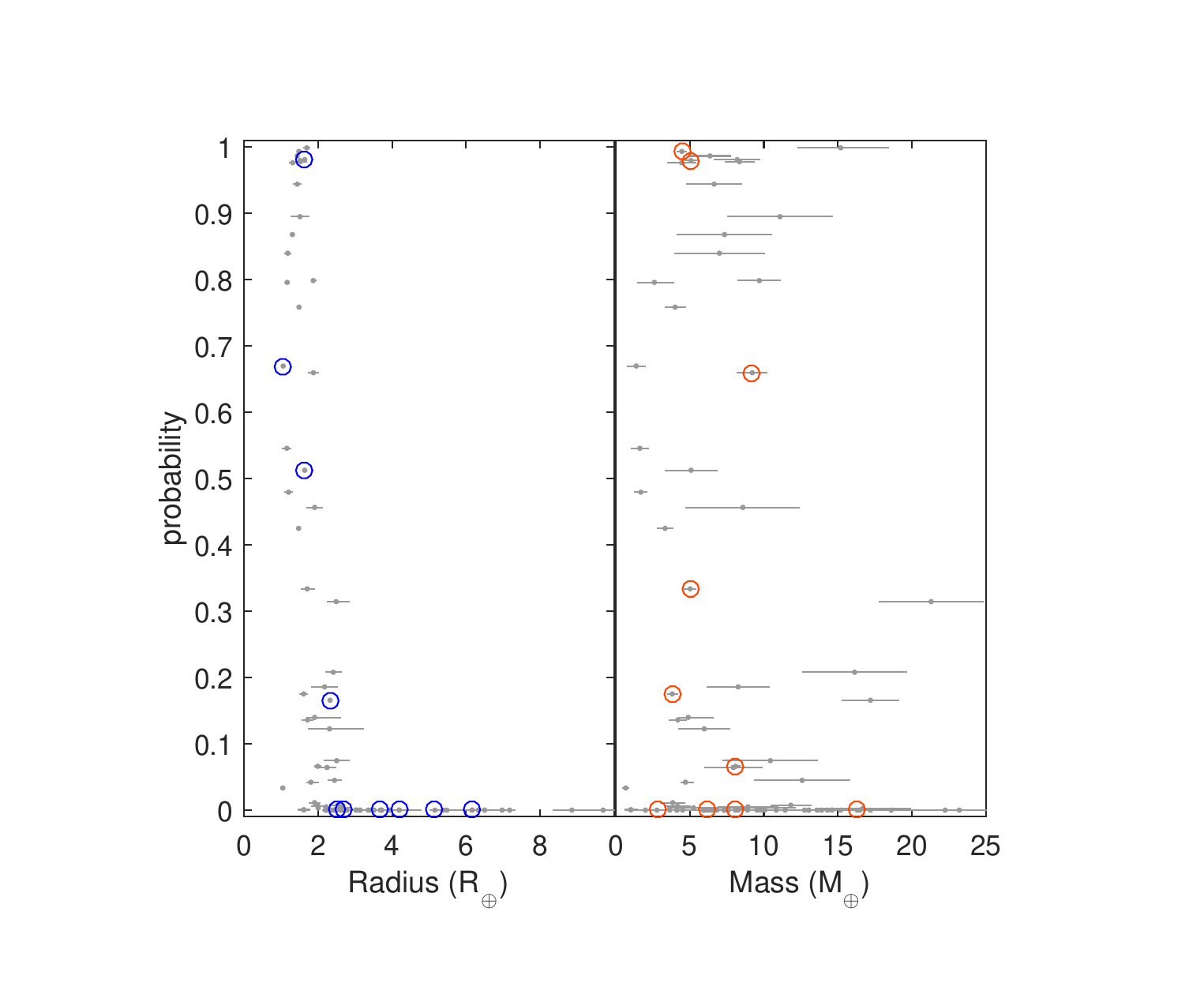}
    \caption{The probability to be denser than pure rock as a function of planetary radius  (left) and mass (right). 
    The ten planets with the best radius or mass determination are highlighted by the blue and orange circles, respectively. While the threshold on the radius is significant, the threshold on a mass is less distinct. }
    \label{fig:MvcR}
\end{figure}

\begin{figure}

	\includegraphics[width=\columnwidth, trim={3.5cm 9.0cm 4.0cm 6.5cm},clip]{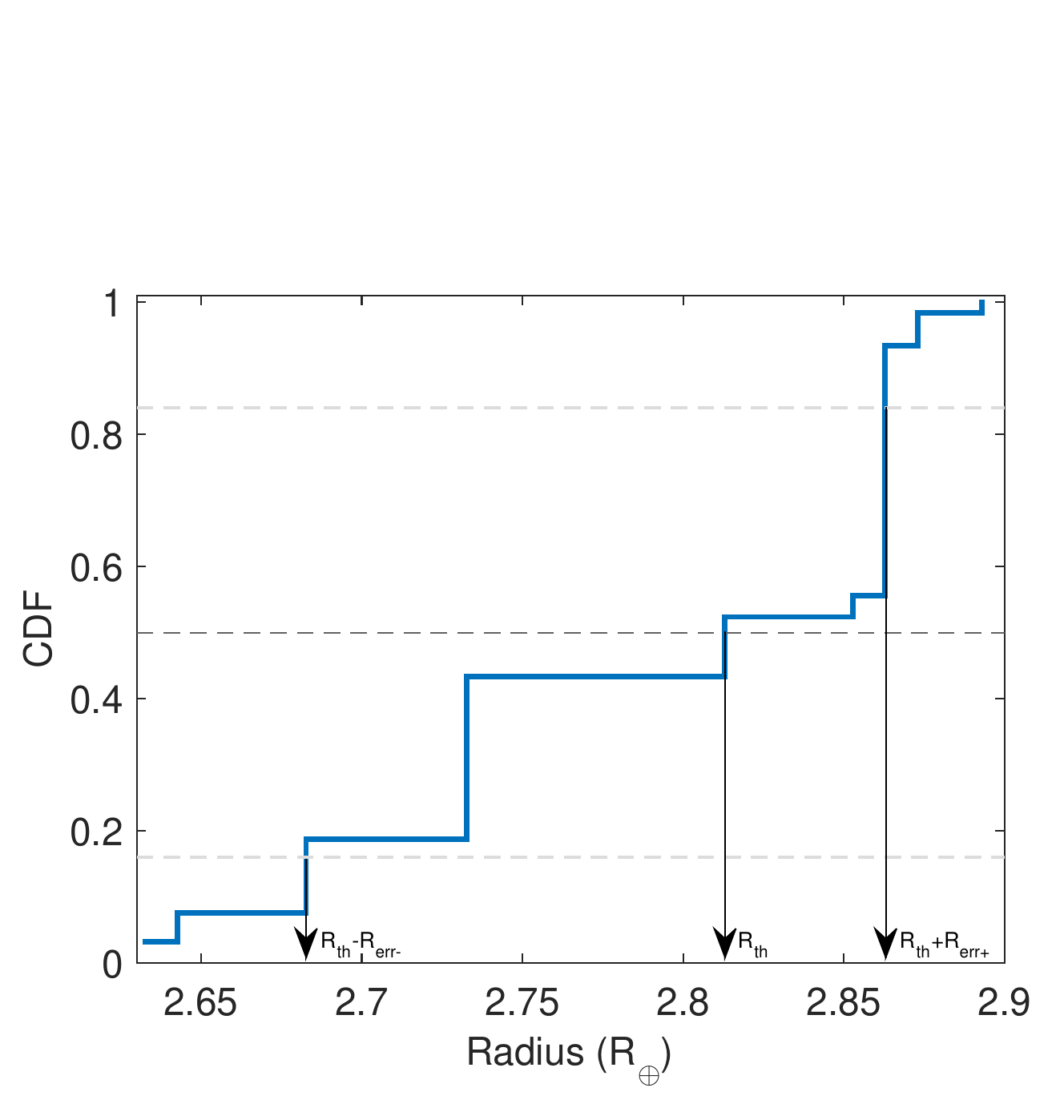}	
    \caption{Cumulative distribution function (CDF) of the R$_{th}$ for a \textit{scenario-1} composition with 2\% of H-He and Z=0.2. The median of R$_{th}$ is evaluated at the 50-th percentile, while the lower ($R_{err-}$) and upper ($R_{err+}$) uncertainty values  are taken to be the 16-th and 84-th percentiles. }
    \label{fig:cdf}
\end{figure}

We investigate whether we can infer threshold masses $M_{th}$ using a similar analysis as the one used for the radii. 
We perform the test for the case of a pure-rock composition which is very robust. 
The results are presented in Figure \ref{fig:MvcR}. 
Unlike the radii (left panel), while there is a similar trend for the masses (right panel), the existence of $M_{th}$ is less conclusive. 
In the figure, we highlight ten planets with the most accurate radius/mass determination. Also for this sample, the threshold on a mass is less distinct.

The inferred distributions of $p$ as a function of planetary mass for all the compositions we considered (not shown) do not resemble a  sharp step-function of a form of Equation \ref{eq:step}. 
Reducing the sample to ten planets with the smallest relative measurement uncertainties (orange circles) leads to a similar conclusion. 
This suggests that the distribution of masses is more continues than that of the radii and that there are no sharp mass boundaries. 
Nevertheless, at the moment, we cannot exclude the possibility that threshold masses do not exist. In order to do so it is desirable to have a large number of planets with accurate measured masses. 
At the moment, we can only conclude that our sample of planets cannot be used to determine threshold masses using the same methodology. Accurate measurements with $\sim 5\%$ uncertainty on {\it both} the planetary mass and radius as expected by PLATO with the masses being determined via ground-based radial-velocity followups \citep{Rauer2014}. This could reveal the existence of $M_{th}$ and can be used to further refine the threshold radii. 

\subsubsection{The photoevaporation valley}
%
%
The bimodal size distribution of the \textit{Kepler} planets reported by \citet{Fulton2017, Fulton2018} and explained by photoevaporation models \citep{Owen2017,Lopez2014,Jin2014}, suggests that most exoplanets 
originally formed with H-He,  but the less massive planets lost it at a later stage due to their low gravity and strong irradiance from the host star. 
Since the valley falls in the size-range of 1.5-2 \RE, they adopt a radius of 1.7 \RE \, to delimit between Super-Earths and mini-Neptunes. Thus, under the photoevaporation interpretation, the definitions of super-Earth and mini-Neptune do not reflect a difference in the origin of the objects, but on their evolution.
In our study, we provide threshold radii based on the existing data, and therefore cannot provide predictions that go beyond the available data.  
%
Our method does not exclude the possibility that some planets with radii of $\sim$ 2 R$_\oplus$ have H-He atmospheres, as is inferred from the work of Fulton. 
It is interesting to note that the gap of planets in the size-range of 1.5-2 \RE \, is not empty \citep{Fulton2018}. It is hard to reconcile this with a scenario where planets are composed of a pure rocky-core surrounded solely by a H-He envelope: planets with radius in the gap should be unstable towards photoevaporation, losing H-He until reaching the first peak of the distribution \citep{Owen2017}. Perhaps this suggests that exoplanetary atmospheres are typically enriched with heavy elements and are not made of pure H-He. 

%

\section{Summary and Conclusions}
\label{sec:Conclusions}

We present a statistical analysis to determine the threshold radii of volatile-rich planets. 
We show that different assumed compositions and internal structures with fixed $\fHHe$ and $Z$ lead to a range threshold radii. 
As a result, in order to characterize individual planets information on their orbital properties and atmospheric compositions is required.  
However, despite the degenerate nature of the problem we suggest that there are characteristic threshold radii for different compositions. 
\par

First, we confirm that planets with radii larger than 1.6 R$_\oplus$ are not rocky, and must consist of lighter elements. 
This conclusion is consistent with the work of \citet{Rogers2015}, despite some differences in the statistical analysis and the used planetary sample.  
It is found that distinguishing a pure-water planet from a rocky planet with a thin H-He atmosphere is not possible. 
Therefore, planets that are classified as ocean planets might in reality be rocky core planets with a volatile atmosphere \citep{Adams2008}. 
\par

Second, we show that most of the planets larger than $\sim 3$  R$_\oplus$ must contain at least 2\% of H-He, 
	while most of the planets with radii larger than  $\sim 3.6$ R$_\oplus$ and 4.3 R$_\oplus$ must contain at least 5\% and 10\% of H-He, respectively. 

While the exact value of R$_{th}$ depends on the model assumptions (i.e., composition, structure, thermal state, EOS), we find a range of threshold radii of $\sim 2.5-4.3 R_\oplus$ for planets with rocky cores and gaseous atmospheres. 
These radii are typically larger than the threshold radii for pure-water planets ($R_{th}$ $\sim 2.6 R_\oplus$). 
We find that although the planetary albedo and semi-major axis affect the planetary temperature, they have a relatively small impact on the inferred R$_{th}$. 
For albedos between zero and 0.9, R$_{th}$ varies from $\sim$ 3 $R_\oplus$ to $\sim$ 3.2 $R_\oplus$, suggesting that assumed albedo has a very small impact on R$_{th}$. 
We suggest that high planetary luminosity leads to somewhat larger R$_{th}$, in the range of sensible luminosities ($L \sim 10^{-1}-10^{2} L_N $, where $L_N$ is Neptune's luminosity) the change in R$_{th}$ is very small, suggesting that R$_{th}$  is relatively insensitive to the assumed planetary luminosity. 
\par

%
The key conclusions of our study can be summarized as follows: 
\begin{itemize}
	\sitem We confirm that planets with radii larger than $\sim$1.6 R$_\oplus$ are not pure-rocky worlds and must consist of lighter elements. 
	\sitem Planets with radii larger than $\sim$2.6 R$_\oplus$ are not pure-water worlds and must consist of atmospheres (presumably of H-He). 
\sitem By defining a mini-Neptune (or a Neptune-analog) as a planet with at least 2\% of H-He in mass, we find that  the transition from super-Earths (planets consisting of less than 2\% of H-He) to mini-Neptunes occurs at $\sim$ 3 R$_{\oplus}$.
	\sitem Planets with radii larger than $\sim$4 R$_\oplus$ are expected to consist of at least 10\% of H-He and are therefore gaseous-rich. 
\end{itemize}

Upcoming data from space missions such as CHEOPS, TESS and PLATO as well as ground-based facilities will further constrain the possible compositions of exoplanets. 
Measurements of planets with similar masses at larger radial distances will allow us to extend our scheme and characterize colder planets and reveal whether the threshold radii are expected to change with the distance to the host star. Finally, accurate measurements of both the masses and radii of small- and intermediate- mass exoplanets will allow us to determine whether threshold masses exist. This will significantly improve our understanding of the formation, and evolution, and internal structures of planets in the solar neighbourhood. 

\newpage
\bibliography{mybib} 

\section*{Acknowledgements}
M.L. thanks Omri Harosh and Uzi Vishne for valuable remarks and advices on the statistical analysis. R.H. acknowledges support from SNSF grant 200021$\_$169054. Some of this work has been carried out within the framework of the National Centre for Competence in Research PlanetS, supported by the Swiss National Foundation.

\appendix
\section{Atmospheric model}
\label{AtmMdl}

The mass-radius relations are derived using the standard structure equations of hydrostatic equilibrium, mass conservation, and heat transport for the gaseous envelope:

\begin{subequations}
\begin{equation}
	{\mbox{d} P \over \mbox{d} r} = - { G m \rho \over r^2 },
\end{equation}
\begin{equation}
	{\mbox {d} m \over \mbox{d} r} = 4 \pi r^2 \rho ,
\end{equation}
\begin{equation}
	{\mbox {d} T \over \mbox{d} r} = \frac{T}{P}\frac{\mbox {d} P}{\mbox {d}  r} \nabla,
\end{equation}
\end{subequations}
where $r$ is the radius, $m$ is corresponding cumulative mass, $\rho$ is a density of each shell, $P$ is a pressure, $G$ is the gravitational constant,  $\sigma$ is Stephan-Boltzmann constant, $L$ the intrinsic luminosity, and $\nabla$ the minumum between the adiabatic  and radiative gradient (Eq.\ref{RadGrad}).


To account for irradiation, we use a semi-gray atmosphere model \citep{Guillot2010,Jin2014}, in which two opacity sources are included: the visible ($\kappa_v$) and infrared ($\kappa_{th}$) mean opacities. 
The optical depth is computed, which is given by:
\begin{equation}
\frac{d \tau}{dr} = \kappa_{\text{th}} \rho
\end{equation}
being $\kappa_{\text{th}}$ the infrared mean opacity, taken as $\kappa_{\text{th}} = 0.01$ g/cm$^3$  \citep{Guillot2010}. 

The temperature gradient of the irradiated atmosphere is given by \citep{Guillot2010}:

\begin{equation}\label{guillot}
\begin{aligned}
T^4 = & \frac{3 T_{\text{int}}^4 }{4} \bigg[\frac{2}{3} + \tau \bigg] + \frac{3 T_{\text{eq}}^4 }{4} \bigg[\frac{2}{3} + \frac{2}{3 \gamma}\bigg\{1 + \bigg(\frac{\gamma \tau}{2} -1\bigg) e^{-\gamma \tau} \bigg\} \\ 
& + \frac{2\gamma}{3} \bigg( 1 - \frac{\tau^2}{2} \bigg) E_2(\gamma \tau) \bigg]
\end{aligned}
\end {equation}
where $\gamma = \kappa_{\text{v}}/ \kappa_{\text{th}}$ (ratio between visible and infrared opacity), $T_{\text{int}}$ is the intrinsic temperature given by $T_{\text{int}} = (L/(4\pi \sigma r^2))^{1/4}$, and $E_{2} (\gamma \tau)$ is the exponential integral, defined by $E_{n}(z) \equiv \int_1^{\infty} {t^{-n} e^{-zt} dt}$ with $n = 2$. $\gamma$ is taken from the calibration of \citet{Jin2014}.
The boundary between the irradiated atmosphere and the envelope is set at $\gamma \tau = 100 / \sqrt(3)$ \citep{Jin2014}. For $\gamma \tau$ larger than this, the usual Schwarzschild criterion to distinguish between convective and radiative layers is applied. That is, if the adiabatic temperature gradient is larger than the radiative one, the layer is stable against convection, and the radiative diffusion approximation is used for computing the temperature gradient:

\begin{equation}\label{RadGrad}
\frac{dT}{dr} = - \frac{3 \kappa_{\text{th}} L \rho}{64 \pi \sigma  T^3 r^2}
\end{equation}
where $L$ is the intrinsic luminosity. 

The planetary luminosity uses the luminosity fit of \citet{Rogers2010} which corresponds to planet evolution calculations derived by \citet{Baraffe2008}, and is given by: 
\begin{equation}
	\log{\frac{L}{L_{\odot}}}=a_1 + a_{Mp}\log\frac{M_p}{M_\oplus}+a_{Rp} \log\frac{R_p}{R_{jup}}+a_{tp} \log\frac{t_p}{1 Gyr},
		\label{eq:Lumi}
\end{equation}
where $L_{\odot}$ is the solar luminosity, $M_p$ is the planetary mass in Earth masses, $R_p$ is the planetary radius in Jupiter radii, and $t_p$ is the stellar age (taken  to be 5 Gyr). The coefficients are $a_1=-12.46, a_{Mp} =1.74, a_{Rp}=-0.94, a_{tp}=-1.04$.  It should be noted that the atmospheric temperature does not only depend on the intrinsic luminosity alone, but also on stellar irradiation. 

For the fully mixed models ({\it scenario-1}) we assume that the water is homogeneously distributed in H-He in a vapour phase. In {\it scenario-2}, where the planet is assumed to be differentiated, with any liquid water assumed to be isothermal while ice is assumed adiabatic \citep[e.g.][]{Dorn2017}. 
The stellar luminosity is assumed to be solar and the semi-major axis 0.1 AU, corresponding to a temperature of $\sim$ 770 K. 
The sensitivity of the inferred M-R relation to these assumptions is invested in sections \ref{sec:AdjT} and \ref{subsec:Lum}.  
Further details on the structure models can be found in \citet{Venturini2015,Dorn2017} and references therein. 


\newpage
\begin{table}
\centering
\caption{The derived threshold radii R$_{th}$ (as defined in section \ref{subsubsec:Threshold}) and their uncertainties for various possible compositions without volatiles.}
\label{tab:Rocky}
\begin{tabular}{|c  |  c  |}
\hline
{\bf Composition} & {\bf  R$_{th}$} (R$_\oplus$)  \\ \hline
Earth-like  & 1.47$^{+0.11}_{-0.01}$          \\ \hline
Pure-rock        & 1.66$^{+0.01}_{-0.08}$          \\ \hline
Pure-water         & 2.58$^{+0.05}_{-0.05}$          \\ \hline
\end{tabular}
\end{table}

\begin{table}
\centering
\caption{The derived threshold radii $R_{th}$ (in R$_\oplus$) assuming different envelope metallicities ($Z$) and H-He mass fractions ($\fHHe$), as defined in section \ref{subsubsec:Key}. Listed are the results for the two structure scenarios of a fully mixed envelope ({\it scenario-1}) and a fully differentiated ({\it scenario-2)} planet (see section \ref{subsec:Mass-Radius Relation} for details). The planetary albedo is assumed to be A=0 and the semi-major 0.1 AU. In some cases the statistical test is inconclusive and cannot provide an estimate of the threshold boundaries. }
\label{tab:VarZvarF}
\begin{tabular}{l|l|l|l|l|l|l|}
\cline{2-7}
                              & \multicolumn{2}{l|}{$\fHHe=2\%$} & \multicolumn{2}{l|}{$\fHHe=5\%$} & \multicolumn{2}{l|}{$\fHHe=10\%$} \\ \cline{2-7} 
                              & mixed     & differentiated     & mixed     & differentiated     & mixed     & differentiated      \\ \hline
\multicolumn{1}{|l|}{Z = 0.0} &2.90$^{+0.02}_{-0.01}$&2.90$^{+0.02}_{-0.01}$&3.49$^{+n/a}_{-0.22}$&3.49$^{+0.04}_{-0.13}$&4.04$^{+0.05}_{-0.00}$&4.06$^{+0.03}_{-0.02}$\\ \hline
\multicolumn{1}{|l|}{Z = 0.1} &2.90$^{+0.02}_{-0.03}$&2.90$^{+0.02}_{-0.01}$&3.31$^{+0.02}_{-0.02}$&3.53$^{+0.03}_{-0.17}$&3.86$^{+0.18}_{-0.01}$&4.06$^{+0.03}_{-0.02}$ \\ \hline
\multicolumn{1}{|l|}{Z = 0.2} &2.81$^{+0.05}_{-0.13}$&2.92$^{+n/a}_{-0.03}$&3.22$^{+0.02}_{-0.01}$&3.53$^{+0.03}_{-0.17}$&3.85$^{+0.01}_{-0.08}$&4.09$^{+n/a}_{-0.03}$\\ \hline
\multicolumn{1}{|l|}{Z = 0.4} &2.59$^{+0.04}_{-0.03}$&2.92$^{+n/a}_{-0.03}$&2.94$^{+0.03}_{-0.02}$&3.53$^{+0.03}_{-0.04}$&3.56$^{+0.01}_{-0.03}$ &4.33$^{+n/a}_{-n/a}$\\ \hline
\multicolumn{1}{|l|}{Z = 0.7} &2.52$^{+0.03}_{-0.14}$&2.97$^{+0.22}_{-0.08}$&2.78$^{+0.12}_{-0.10}$&3.65$^{+n/a}_{-0.12}$&3.19$^{+n/a}_{-0.20}$&4.33$^{+n/a}_{-n/a}$\\ \hline
\end{tabular}
\end{table}

\begin{table}[h]
\centering
\caption{The derived threshold radii for pure-water planets using different water EOSs. }
\label{tab:H2OcompEOS}
  \begin{tabular}{|c|c|}
  \hline
    {\bf H$_2$O EOS }& {\bf  R$_{th}$ (R$_\oplus$) }\\ \hline
    Seiger et. al. 2007  & 2.58$^{+0.05}_{-0.05}$\\ \hline
    Wagner et. al. 2010  & 2.53$^{+0.03}_{-0.08}$\\ \hline		
    More et. al. 1988 & 2.63$^{+0.10}_{-0.05}$\\ \hline			
  \end{tabular}
\end{table}

\begin{table}
\centering
\caption{The derived threshold radii  R$_{th}$ and their uncertainties assuming different semi-major axes and albedos. $\fHHe$ is set to be 5\% with a metallicity of Z=0.2. The planetary envelope is assumed to be fully mixed (\it {scenario-1}). }
\label{tab:Summary3}
\begin{tabular}{|c|c|c|}
\hline
{\bf Semi-Major Axis} (AU) &{\bf Planetary Temperature, $T_{eq}$} (K) & {\bf  R$_{th}$ (R$_\oplus$)} \\ \hline
0.05 &1120           & 3.48$^{+0.03}_{-0.17}$          \\ \hline
0.1 &770             & 3.22$^{+0.02}_{-0.01}$          \\ \hline
0.5 &330	           & 2.92$^{+0.18}_{-0.03}$         \\ \hline
Calculated individually, A=0.0 & Calculated individually    & 3.19$^{+0.08}_{-0.22}$          \\ \hline
Calculated individually, A=0.3 & Calculated individually    & 3.19$^{+0.08}_{-0.22}$          \\ \hline
Calculated individually, A=0.9 & Calculated individually    & 2.97$^{+0.02}_{-0.08}$          \\ \hline

\end{tabular}
\end{table}

\begin{table}[h]
\centering
\caption{The derived threshold radii  R$_{th}$ using different  luminosities (in units of Neptune's luminosity). We present a fully mixed case ({\it scenario-1}) of $\fHHe=5\%$ and $Z=0.20$ with a luminosity set by  equation \ref{eq:Lumi} as described in the Appendix, versus a range of constant luminosities. }
\label{tab:varL}
\begin{tabular}{|c|c|}
 \hline
{\bf Luminosity (L$_N$)} & {\bf R$_{th}$  (R$_\oplus$)} \\ \hline
Standard case &    3.22$^{+0.02}_{-0.01}$\\ \hline
$\sim 10^{-1}$ & $3.19^{+0.01}_{-0.22}$ \\ \hline
$\sim 10^{2}$ & 3.27$^{+0.01}_{-0.08}$ \\ \hline
$\sim 10^{3}$ & $3.53^{+0.03}_{-0.04}$    \\ \hline
\end{tabular}
\end{table}

\begin{table}
\caption{A summary of a data stability test, described at section \ref{subsec:Stability test}. The results compare R$_{th}$ found from the real data and slightly randomized data-sets (modified data). 
The composition used for the testing is {\it scenario-1} with $\fHHe$= 5\% and Z=0.2}
\label{tab:SumStab}
\centering
  \begin{tabular}{|c|c|}
  \hline
    {\bf Planet Sample }& {\bf  R$_{th}$ (R$_\oplus$) }\\ \hline
    Original data & 3.24$^{+0.03}_{-0.03}$\\ \hline
   Modified data sample 1 & 3.22$^{+0.70}_{-0.38}$ \\ \hline
    Modified data sample 2 & 3.22$^{+0.70}_{-0.38}$  \\  \hline
		Modified data sample 3  & 3.31$^{+0.61}_{-0.29}$ \\  \hline

  \end{tabular}
\end{table}

\newpage

\end{document}